\documentclass[twocolumn,conference]{IEEEtran}
\usepackage[T1]{fontenc}
\usepackage[latin9]{inputenc}
\usepackage{color}
\usepackage{array}
\usepackage{booktabs}
\usepackage{units}
\usepackage{multirow}
\usepackage{amsmath}
\usepackage{graphicx}

\makeatletter

\newcommand{\noun}[1]{\textsc{#1}}
\providecommand{\tabularnewline}{\\}

\usepackage[caption=false,font=footnotesize]{subfig}
\usepackage{cite}

\@ifundefined{showcaptionsetup}{}{%
 \PassOptionsToPackage{caption=false}{subfig}}
\usepackage{subfig}
\makeatother

\begin{document}
\title{An Interpretable Neural Network for Configuring Programmable Wireless
Environments}
\author{Christos Liaskos\IEEEauthorrefmark{1}, Ageliki Tsioliaridou\IEEEauthorrefmark{1},
Shuai Nie\IEEEauthorrefmark{3}, Andreas Pitsillides\IEEEauthorrefmark{2},
Sotiris Ioannidis\IEEEauthorrefmark{1}, and Ian Akyildiz\IEEEauthorrefmark{2}\IEEEauthorrefmark{3}\\
{\small{}\IEEEauthorrefmark{1}Foundation for Research and Technology
- Hellas (FORTH), emails: \{cliaskos,atsiolia,sotiris\}@ics.forth.gr}\\
{\small{}\IEEEauthorrefmark{2}University of Cyprus, Computer Science
Department, email: Andreas.Pitsillides@ucy.ac.cy}\\
{\small{}\IEEEauthorrefmark{3}Georgia Institute of Technology, School
of Electrical and Computer Engineering, emails: \{shuainie,ian\}@ece.gatech.edu}}
\maketitle
\begin{abstract}
Software-defined metasurfaces (SDMs) comprise a dense topology of
basic elements called meta-atoms, exerting the highest degree of control
over surface currents among intelligent panel technologies. As such,
they can transform impinging electromagnetic (EM) waves in complex
ways, modifying their direction, power, frequency spectrum, polarity
and phase. A well-defined software interface allows for applying such
functionalities to waves and inter-networking SDMs, while abstracting
the underlying physics. A network of SDMs deployed over objects within
an area, such as a floorplan walls, creates programmable wireless
environments (PWEs) with fully customizable propagation of waves within
them. This work studies the use of machine learning for configuring
such environments to the benefit of users within. The methodology
consists of modeling wireless propagation as a custom, interpretable,
back-propagating neural network, with SDM elements as nodes and their
cross-interactions as links. Following a training period the network
learns the propagation basics of SDMs and configures them to facilitate
the communication of users within their vicinity.
\end{abstract}

\begin{IEEEkeywords}
Wireless, Propagation, Software control, Metasurfaces, Neural Network,
Interpretable.
\end{IEEEkeywords}

\section{Introduction}

With the proliferation of smart wireless devices and burgeons of the
Internet of Things (IoT) over recent years, wireless communication
systems have experienced unprecedented demands for higher data rates,
smaller latency, better services, and lower prices. These requirements
have driven the research efforts into the fifth-generation wireless
networks, narrow-band IoT, among other promising techniques to tackle
the open problems of limited spectrum resources and high user densities~\cite{akyildiz20165g}.
Most of the proposed solutions focus on the improvement of transceivers,
for example, the massive MIMO technique and millimeter wave solutions.
However, a largely overlooked factor that directly influences the
performance of the wireless system is the propagation environments.
With the use of novel materials, the propagation environments can
be turned into \emph{programmable} media, yielding unparalleled, wired-level
gains in wireless power transfer, and the mitigation of interference,
Doppler effects and malicious eavesdropping~\cite{liaskos2018new}.

\begin{figure*}[!t]
\centering{}\textcolor{black}{\includegraphics[viewport=0bp 0bp 880bp 400bp,clip,width=0.65\textwidth]{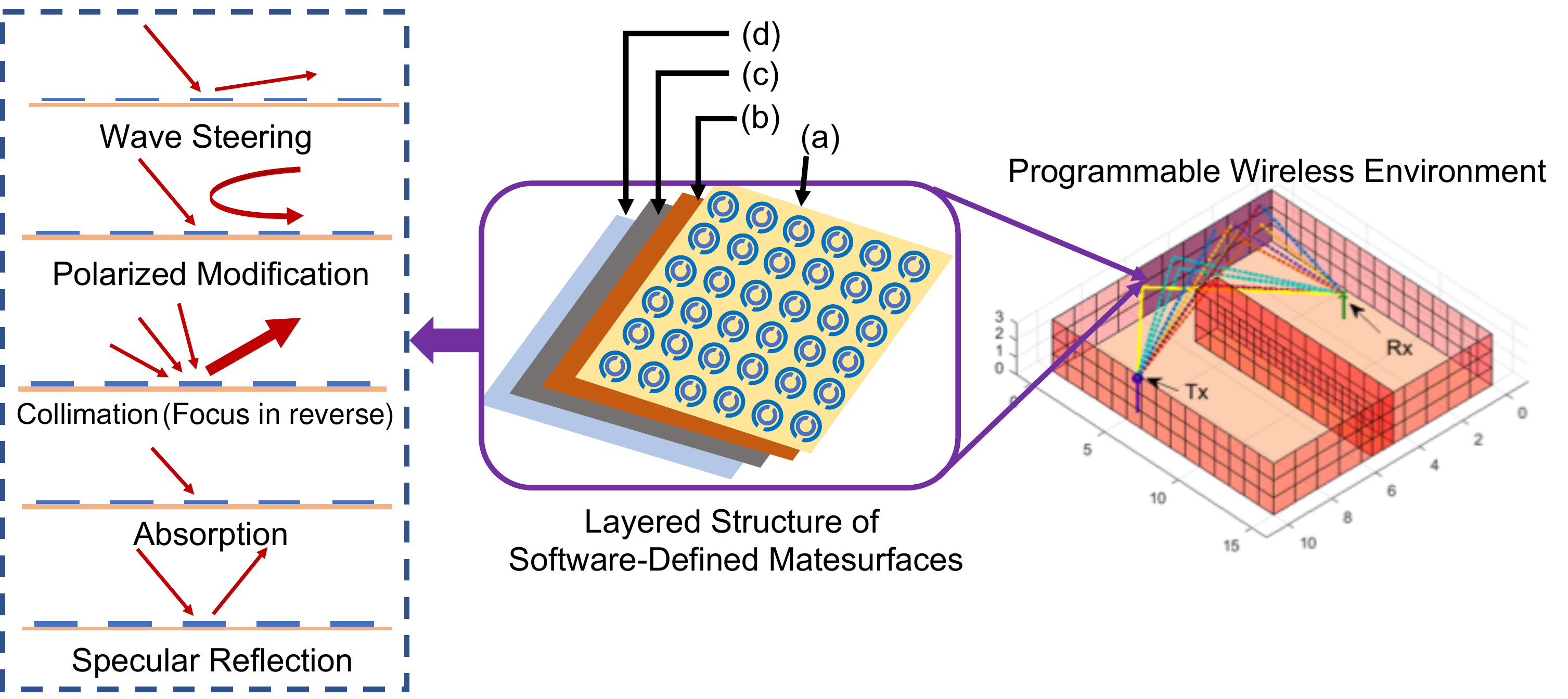}
\caption{\label{fig:structure}Illustration of the SDM tile structure and its
indicative EM functionalities (left), and the enabling operation mode
in a PWE with a pair of transmitter and receiver (right). The layers
of a SDM from top to bottom include (a) the metasurface plane, (b)
the sensing and actuation plane, (c) the computing plane, and (d)
the communications plane.}
}
\end{figure*}

In regular wireless propagation environments, EM waves undergo various
interactions including reflections, diffractions, and scattering.
Rich multi-path conditions are generated as a result of these interactions
which, if not well-controlled, could destructively influence the users'
communication. In order to exert deterministic, adaptive control over
the wireless propagation phenomenon, a novel solution based on artificial
planar materials has been conceived, which utilizes software-defined
metasurfaces (SDMs)\ \cite{liaskos2015design}. SDMs sense and apply
transformative control over EM waves impinging upon them, re-engineering
their direction, polarization and phase, performing arbitrary wavefront
shaping which indicatively includes focus, collimation and absorption
(Fig.\ \ref{fig:structure}-left). SDM \emph{tiles} are rectangular,
connected SDM units that bear an IoT gateway, allowing a central computer
to get or set their desired EM functionality. Such materials have
also been recently generalized under the term \emph{intelligent surfaces}~\cite{Huang2018,wu2018intelligent,wu2019beamforming,hu2018beyond},
to denote their end-facility while abstracting the underlying physical-layer
technology specifics.

Programmable wireless environments (PWEs) are created by coating major
parts of a setting (e.g., ceilings in a floorplan\ \cite{Liaskos2019ADHOC})
with SDM tiles\ \cite{liaskos2018new,Liaskos2018using}, as shown
in Fig.\ \ref{fig:structure}-right. A central computer senses the
EM profile and connectivity objectives of users present in the environment,
and adaptively \emph{configures} the matching functionality for each
SDM tile\ \cite{liaskos2018modeling}.

In this paper, we propose an approach based on machine learning algorithms,
in particular, neural networks, to adaptively configure PWEs for a
set of users. The key-idea is that, since SDM tiles can regulate the
distribution of power within a space, they can be represented by nodes
in a neural network, while the power distribution can be mapped to
neural network links and their weights. The weights are optimized
via custom feed-forward/back-propagation processes, and are interpreted
into SDM tile functionalities. The machine learning approach is shown
to be intuitive in its representation and economic in its user of
SDM tiles, compared to related approaches\ \cite{liaskos2018modeling}.

The remainder of this paper is organized as follows. Section~\ref{sec:background}
provides background information on the principles of SDMs and neural
networks. Section~\textcolor{black}{\ref{sec:NNConfig} details the
proposed neural net approach for configuring PWEs. Evaluation via
ray-tracing follows in Section\ \ref{sec:Evaluation}, along with
a discussion of future directions. The paper is concluded in Section\ \ref{sec:Conclusion}}

\section{Background\label{sec:background}}

\subsubsection*{The structure of SDMs}

A metasurface is an engineered structure which is the two-dimensional
representation of metamaterials with the basic element called ``meta-atom''.
In usual compositions, the substrate is dielectric and the meta-atom
is conductive, which can be made of copper over silicon. Alternative
conductors can be silver, gold, or graphene for operating frequencies
in the terahertz band~\cite{zhu2017traditional,fallahi2012design}.
Metasurfaces can be designed to support the propagation of highly
confined surface plasmon polariton (SPP) waves at microwave, millimeter
wave, and even higher frequencies and to control these EM waves impinging
on them~\cite{lockyear2009microwave}. In general, metasurfaces comprise
several hundreds of meta-atoms, which results into fine-grained control
over the EM wave interactions. In particular, the size of each meta-atom
is comparable to the minimum intended interaction wavelength,~$\lambda$,
with $\frac{\lambda}{10}\to\frac{\lambda}{5}$ constituting a common
choice. The thickness of the metasurfaces is in the sub-wavelength
scale, ranging between $\frac{\lambda}{10}$ and $\frac{\lambda}{5}$
as well.

The layered structure to enable various operation modes of metasurfaces
is given in Fig.~\ref{fig:structure}. The meta-atoms can have different
shapes, including the shown split-ring structure and more complicated
ones~\cite{Liaskos2018using}. The total EM response of the metasurface
is then derived as the total emitted field by all surface currents,
and can take completely engineered forms, achieving custom phase shift,
polarization tuning, and so on. In fact, the meta-atoms can be viewed
as either input or output antennas, connected in custom topologies
via the switch elements. Impinging waves enter from the input antennas,
get routed according to the switch element states, and exit via the
output antennas, exemplary achieving customized reflection.\textcolor{magenta}{}

\subsubsection*{Neural Network for PWEs}

Among many machine learning algorithms, neural networks are known
for their high efficiency in finding optima to complex problems and
making accurate predictions. A neural network consists of an input
layer with input units, at least one layer of hidden nodes neurons,
and an output layer with multiple output units~\cite{Sze.2017}.
The interconnected nodes and layers have associated weights and activation
levels to the links, while each node has an input function, an activation
function, and an output. For a PWE application of neural networks,
the general concept can be described as follows. The input layer units
are configured by the propagation environments including the numbers
and locations of transmitters (TXs) and receivers (RXs), densities
and dimensions of metasurface tiles, operating frequencies, noise
levels, among others. The corresponding output links can be activated
at certain levels. For example, if a metasurface tile is in proximity
to the TX and a non-line-of-sight RX which can be adjusted by its
azimuth angle to reflect the impinging signal towards the RX, the
activation function is related to the received signal strength indicator
(RSSI) and the angle of arrival (AoA) of the signal.

In order for the neural network to generate a desired model, the process
of determining weights to nodes in each layer is of critical importance.
Among many options, backpropagation is the most widely employed method
in weight optimization. The idea of backpropagation is to adjust the
weights of nodes based on the errors at the output layer~\cite{dorner2018deep}.
Since in the neural network each hidden node contributes to the resultant
error of its connected output nodes, the error can be divided according
to the strength of each connection between a hidden node and the output
node, hence the resultant error is propagated back to each layer to
adjust the weights which will eventually minimize the new error.

It is noted that the present work employs custom neural networks,
going beyond their standard composure and conceptual operation. We
proceed to detail the specifics of the proposed solution in the next
Section.

\section{\noun{NNConfig}: Neural Net-Configured PWEs\label{sec:NNConfig}}

\begin{figure}[t]
\begin{centering}
\includegraphics[width=1\columnwidth]{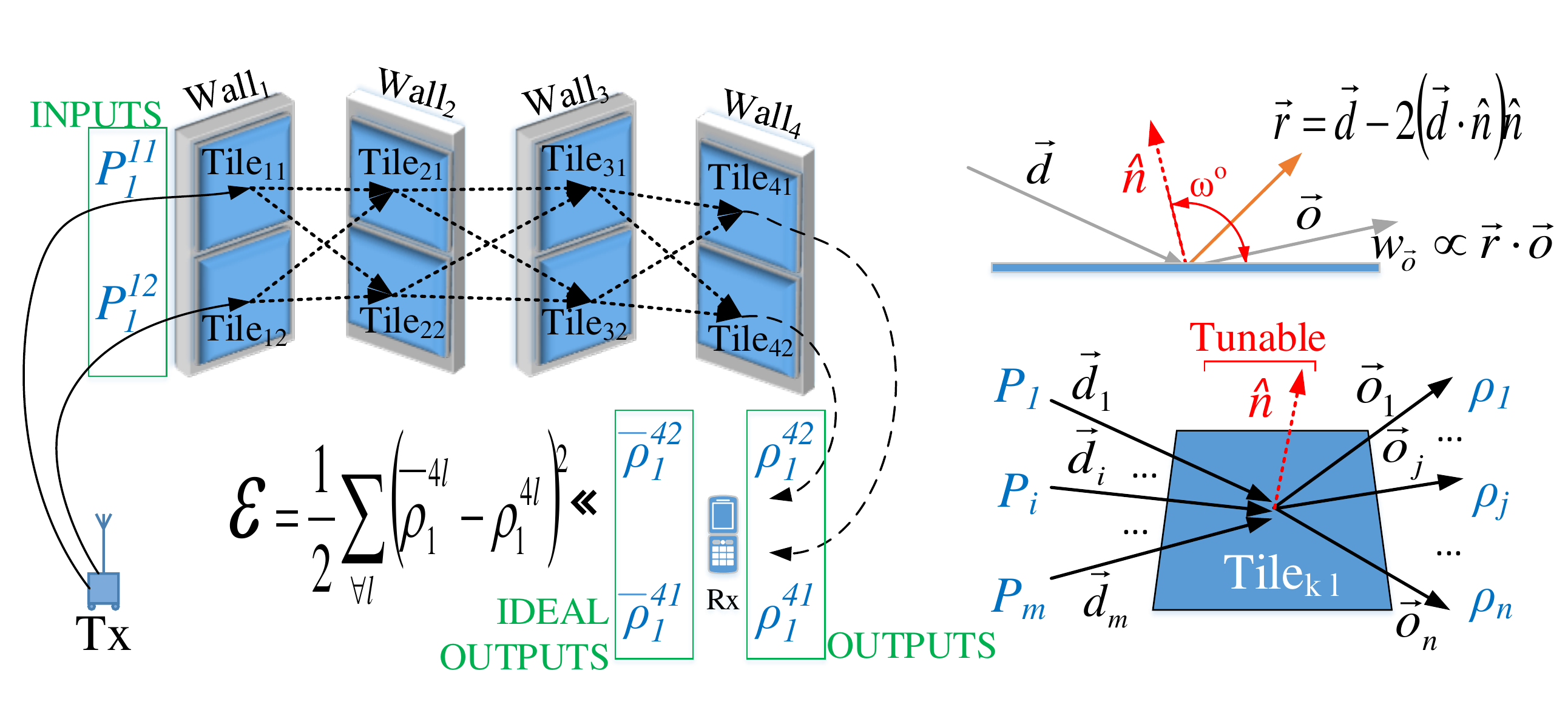}
\par\end{centering}
\caption{\label{fig:concept}Conceptual representation of wireless propagation
as a neural network. }
\end{figure}
Consider a PWE environment comprising a wireless transmitter (Tx),
a receiver (Rx) and a set of walls-scatterers coated with SDM tiles.
This environment is illustrated in Fig.~\ref{fig:concept}.

We assume that EM waves transmitted from the Tx impinge upon the first
wall. Thus, each SDM tile over wall$_{1}$ has its own ``input''
impinging power. Each tile can be tuned to split and redirect its
impinging power to other walls and tiles at their line-of-sight (LOS).
These LOS links are shown as dashed arrows in Fig~\ref{fig:concept}.
Finally, the Rx receives part of the originally transmitted power
via tiles in his LOS. The received power from each Rx link can be
considered as the ``output'' of the propagation process. The ``ideal
output'' represents lossless propagation, i.e., receiving the total
of the transmitted power. The distribution over the Rx links respects
the corresponding receiving gains of the Rx device, as derived by
the Rx antenna pattern and the active MIMO configuration. The ideal
and received outputs can be compared, in order to obtain a metric
of deviation, $\mathcal{E}$, as shown in Fig.~\ref{fig:concept}.

We proceed to study the inputs/outputs of a single tile and their
relation. Following the SDM model of~\cite{liaskos2018modeling},
the SDM tile steering can be modeled as virtually rotating the tile
surface normal unit vector $\hat{n}$. The steering effect is shown
in the top-right inset of Fig.~\ref{fig:concept}. The reflected
wave direction, $\vec{r}\left(\hat{n},\overrightarrow{d}\right)=\overrightarrow{d}-2\left(\overrightarrow{d}\cdot\hat{n}\right)\hat{n}$~\cite{hitzer2013introduction},
for a wave impinging from a direction $\overrightarrow{d}$ can be
altered by tuning $\hat{n}$, which is achieved by setting the SDM
active elements accordingly. This information is provided by the SDM
manufacturer, e.g., in the form of a lookup table.

A reflection $\vec{r}\left(\hat{n},\overrightarrow{d}\right)$ may
not coincide exactly with a given inter-tile, ``outgoing'' link
direction $\overrightarrow{o}$. Thus, we proceed to make a\emph{
heuristic} assumption that the power fraction, $w_{\overrightarrow{o}}$,
over an ``outgoing'' link $\overrightarrow{o}$ is:
\begin{equation}
w_{\overrightarrow{o}}\left(\hat{n},\overrightarrow{d}\right)=\frac{\max\left\{ \vec{r}\left(\hat{n},\overrightarrow{d}\right)\cdot\vec{o},\,0\right\} }{\sum_{\forall\overrightarrow{o}}w_{\overrightarrow{o}}},\label{eq:w}
\end{equation}
i.e., the projection of $\vec{r}$ over $\vec{o}$ normalized across
all outgoing links, assuming that $\vec{o}$ and $\vec{d}$ are unit
vectors. A negative inner product in relation~(\ref{eq:w}) implies
that vectors $\vec{r},\,\vec{o}$ do not face in the same direction
and, thus, $w_{\overrightarrow{o}}$ should intuitively be zero.

Based on these remarks, we proceed to model the wireless propagation
as a neural network with walls as layers and tiles as nodes, and define
feed-forward and back-propagate processes as follows:

\subsubsection{Feed-forward}

Consider the general case of tile$_{k,l}$, indexed and denoted as
shown in Fig.~\ref{fig:concept}. The power at each of its outgoing
links is:
\begin{equation}
\rho_{j}^{\left(k,l\right)}=\sum_{\forall\overrightarrow{d}_{i}}w_{\overrightarrow{o}}\left(\hat{n},\overrightarrow{d_{i}}\right)\cdot P_{i}^{\left(k,l\right)},
\end{equation}
which serves as input power for the next wall-layer, since SDMs can
employ collimation functionalities to focus their total outgoing power
to the next tile~\cite{liaskos2018modeling}. At the final wall-layer,
$k=\kappa$, the metric of deviation, $\mathcal{E}$, is calculated
as:
\begin{equation}
\mathcal{E}=\frac{1}{2}\sum_{\forall l}\left(\overline{\rho}_{1}^{\left(\kappa,l\right)}-\rho_{1}^{\left(\kappa,l\right)}\right)^{2}=\frac{1}{2}\sum_{\forall l}\delta_{l}^{2},
\end{equation}
where $\overline{\rho}$ denotes ideal output power ratio, and $\delta_{l}=\overline{\rho}_{1}^{\left(\kappa,l\right)}-\rho_{1}^{\left(\kappa,l\right)}$.

\subsubsection{Back-propagation}

Once the walls-layers have been iterated over until the final layer,
each tile-node at each wall-layer (in reverse order) deduces the effect
of its currently active $\hat{n}$ vector to the deviation $\mathcal{E}$,
and updates $\hat{n}$ to a new value $\hat{n}_{*}$. For simplicity
we will focus on the 2D case where $\hat{n}$ is defined by a single
angle $\omega$, as shown in Fig.~\ref{fig:concept} (top-right).
The corresponding update rule from $\omega$ to $\omega_{*}$ at tile
$\left(k,l\right)$ follows the generalized delta rule~\cite{DeltaRule}:
\begin{equation}
\omega_{*}^{\left(k,l\right)}=\omega^{\left(k,l\right)}-\eta\cdot\left(\frac{\partial\mathcal{E}}{\partial\omega}\right)^{\left(k,l\right)}\cdot\mathcal{S}^{\left(k,l\right)},
\end{equation}
where $\eta\in\left(0,1\right]$ is the network's learning rate, and
$\mathcal{S}^{\left(k,l\right)}$ is a factor denoting the significance
of tile $\left(k,l\right)$ with regard to the propagation. For the
final wall-layer $k=\kappa$ we define that $\mathcal{S}^{\left(\kappa,l\right)}=\delta_{l}$
(i.e., its deviation from the local ideal output). For $k\ne\kappa$
we define it as the total power impinging on the tile, i.e., $\mathcal{S}^{\left(\kappa,l\right)}=\sum_{\forall i}P_{i}^{\left(k,l\right)}$.

Finally, we provide the following formulas for $\left(\frac{\partial\mathcal{E}}{\partial\omega}\right)^{\left(k,l\right)}$
per indexed layer, omitting the proofs:

\begin{equation}
\begin{array}{l}
\left(\frac{\partial\mathcal{E}}{\partial\omega}\right)\stackrel{\left(\kappa,l\right)}{=}\delta_{l}\cdot\frac{\partial\rho_{1}^{\left(\kappa,l\right)}}{\partial\omega^{\left(\kappa,l\right)}}\\
\left(\frac{\partial\mathcal{E}}{\partial\omega}\right)\stackrel{\left(\kappa-1,l\right)}{=}\underset{\forall j}{\sum}\underset{e_{j}}{\underbrace{\frac{\partial\rho_{j}^{\left(\kappa-1,l\right)}}{\partial\omega^{\left(\kappa-1,l\right)}}}}\cdot\underset{a_{j}}{\underbrace{\delta_{j}\cdot w_{\overrightarrow{o_{j}}}^{\left(\kappa-1,l\right)}}}=\mathbf{e^{\left(\kappa-1,l\right)}\cdot\mathbf{a}^{\left(\kappa-1,l\right)}}\\
\cdots\\
\left(\frac{\partial\mathcal{E}}{\partial\omega}\right)\stackrel{\left(k,l\right)}{=}\underset{\forall j}{\sum}\frac{\partial\rho_{j}^{\left(k,l\right)}}{\partial\omega^{\left(k,l\right)}}w_{\overrightarrow{o_{j}}}^{\left(k,l\right)}\left(\mathbf{1}\cdot\mathbf{a^{\left(j,l+1\right)}}\right),
\end{array}\label{eq:de}
\end{equation}
where $\mathbf{e}$ and $\mathbf{a}$ are helping vectors defined
as shown above, and $\mathbf{1}$ is an all-ones vector with size
equal to $\mathbf{a}$. $\mathbf{e}$ is only used for facilitating
the definition of $\mathbf{a}$ in relation (\ref{eq:de}) and has
not further use, while $\mathbf{a}$ is updated recursively at each
node. Its elements are the link weight products, for all paths leading
from a right-layer neighbor to any output layer node.

The described feed-forward/back-propagate cycles can be executed in
an online or offline manner, until the deviation $\mathcal{E}$ stabilizes,
reaches an acceptable level or an allocated computational time window
expires. At this point, the PWE controller simply deploys EM functionalities
at each tile $\left(k,l\right)$, matching the attained $\omega^{\left(k,l\right)}$
(and, thus $\hat{n}^{\left(k,l\right)}$) values. This trait makes
the proposed approach a directly interpretable neural network configurator
for PWEs. We denote this proposed approach as \noun{NNConfig }and
proceed to evaluate it via full PWE simulations.

\section{Evaluation\label{sec:Evaluation}}

We evaluate the performance of \noun{NNConfig} in the PWE simulator
presented in~\cite{liaskos2018modeling}. We seek to evaluate the
potential of the proposed scheme in a ray-tracing setting, comparing
its outcomes to: i) regular propagation, and ii) the \noun{KpConfig}
scheme for PWEs presented in~\cite{liaskos2018modeling}.

\noun{KpConfig} is novel scheme for configuring PWEs, which is based
on the ray-routing principle~\cite{liaskos2018modeling}. The multiple
rays that are emitted from a set of transmitters are minutely managed
until their reception from the appropriate receivers, applying proper
EM functions (steering, absorbing, collimating, polarizing, phase-modifying,
etc.) to each tile. Its advantages include versatility in handling
multiple users and multiple objectives, spanning QoS optimization,
Doppler effect mitigation, wireless power transfer and security. One
of the \noun{KpConfig} principles is to use only one function per
tile whenever possible. This implies the management of just one ray
per tile, a principle that can allow for less complex and potentially
more efficient EM functions, but also increases the number of active
tiles.

\begin{table}[t]
\centering{}\caption{\label{tab:Floorplan}Floorplan, user location and beam orientation
coordinate system. The origin is at the lower left corner of the floorplan.}
\begin{tabular}[b]{ccc}
\begin{tabular*}{3.8cm}{@{\extracolsep{\fill}}@{}l}
\multicolumn{1}{l}{\textsc{User: Position, $\alpha$,$\phi$,$\theta$}}\tabularnewline
\midrule
0: {[}2.5, 7.5, 0.5{]}, 40$^{o}$, 0$^{o}$, 0$^{o}$ \tabularnewline
1: {[}7.5, 7.5, 0.5{]}, 40$^{o}$, 0$^{o}$, 180$^{o}$ \tabularnewline
\midrule
\multicolumn{1}{l}{\textsc{Comm. Pair: Objective}}\tabularnewline
\midrule
\textcolor{black}{0$\to$1} : $\text{\textsc{Max. Received Power}}$\tabularnewline
\bottomrule
\end{tabular*} & %
\begin{tabular}{c}
\tabularnewline
\includegraphics[width=0.15\columnwidth]{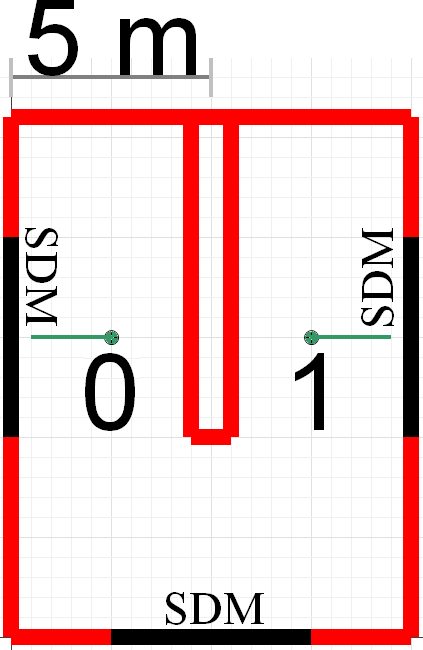}\tabularnewline
\end{tabular} & %
\begin{tabular}{c}
\tabularnewline
\hspace{-15bp}\includegraphics[width=0.33\columnwidth]{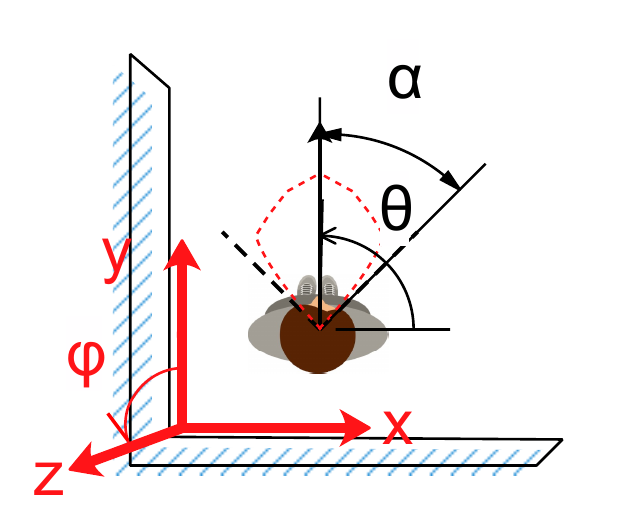}\tabularnewline
\end{tabular} \tabularnewline
\end{tabular}
\end{table}
\begin{table}[t]
\centering{}\caption{\textsc{\label{tab:TSimParams}Simulation parameters.}}
\begin{tabular}{|c|c|}
\hline
Ceiling Height & $1$~$m$\tabularnewline
\hline
Tile Dimensions  & \textbf{$1\times1\,m$} \tabularnewline
\hline
Tile Functions & $\text{\textsc{Steer}}$, $\text{\textsc{Collimate}}$, $\text{\textsc{Absorb}}$\tabularnewline
\hline
Non-SDM surfaces  & Perfect absorbers (cf. red in Table~\ref{tab:Floorplan})\tabularnewline
\hline
Frequency & $2.4\,GHz$ \tabularnewline
\hline
Tx Power & $-30\,dBm$\tabularnewline
\hline
\multirow{2}{*}{Antenna type} & Single $a^{o}$-lobe sinusoid,\tabularnewline
 & pointing at $\phi^{o},\theta^{o}$ (cf. Table~\ref{tab:Floorplan})\tabularnewline
\hline
Max ray bounces & $5$\tabularnewline
\hline
Power loss per bounce & $1$~\%\tabularnewline
\hline
\end{tabular}
\end{table}
We consider a Tx-Rx pair in the floorplan and setup shown in Table~\ref{tab:Floorplan}.
It comprises three walls that the Tx emissions must sequentially and
necessarily hit to reach the Rx. We seek to tune each SDM tile in
this setup so as to maximize the received power at the Rx. Table~\ref{tab:TSimParams}
summarizes the persistent simulator parameters across all subsequent
tests. Note that the setup corresponds to 2D ray propagation in a
3D space (the tile size and floor height are equal), matching the
relations~(\ref{eq:de}) presented in Section~\ref{sec:NNConfig}.

\begin{figure}[t]
\begin{centering}
\subfloat[\label{fig:Untrained}Untrained neural network. ]{\begin{centering}
\includegraphics[width=1\columnwidth]{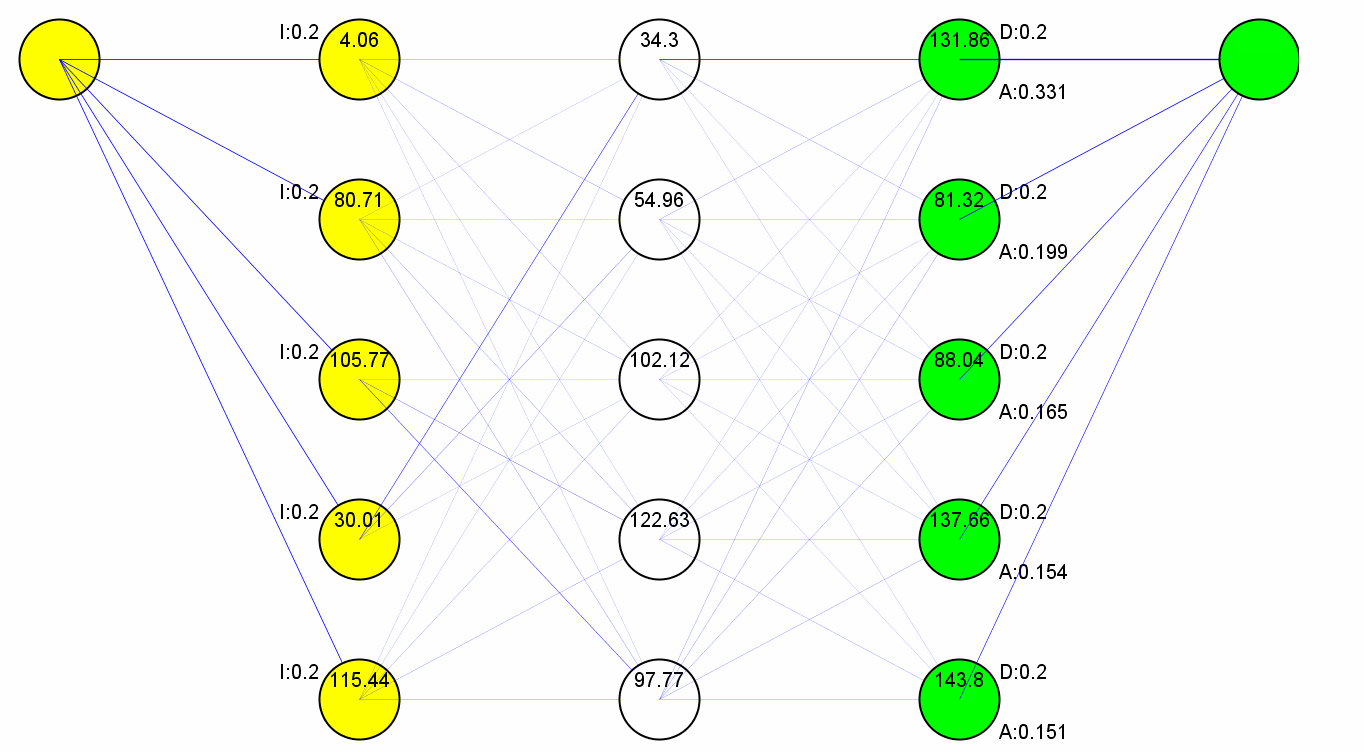}
\par\end{centering}
}
\par\end{centering}
\begin{centering}
\subfloat[\label{fig:Trained}Trained neural network after $2000$ feed-forward/back-propagate
cycles (learning rate $\eta=0.95$, attained RMSE=$8\cdot10^{-4}$). ]{\begin{centering}
\includegraphics[width=1\columnwidth]{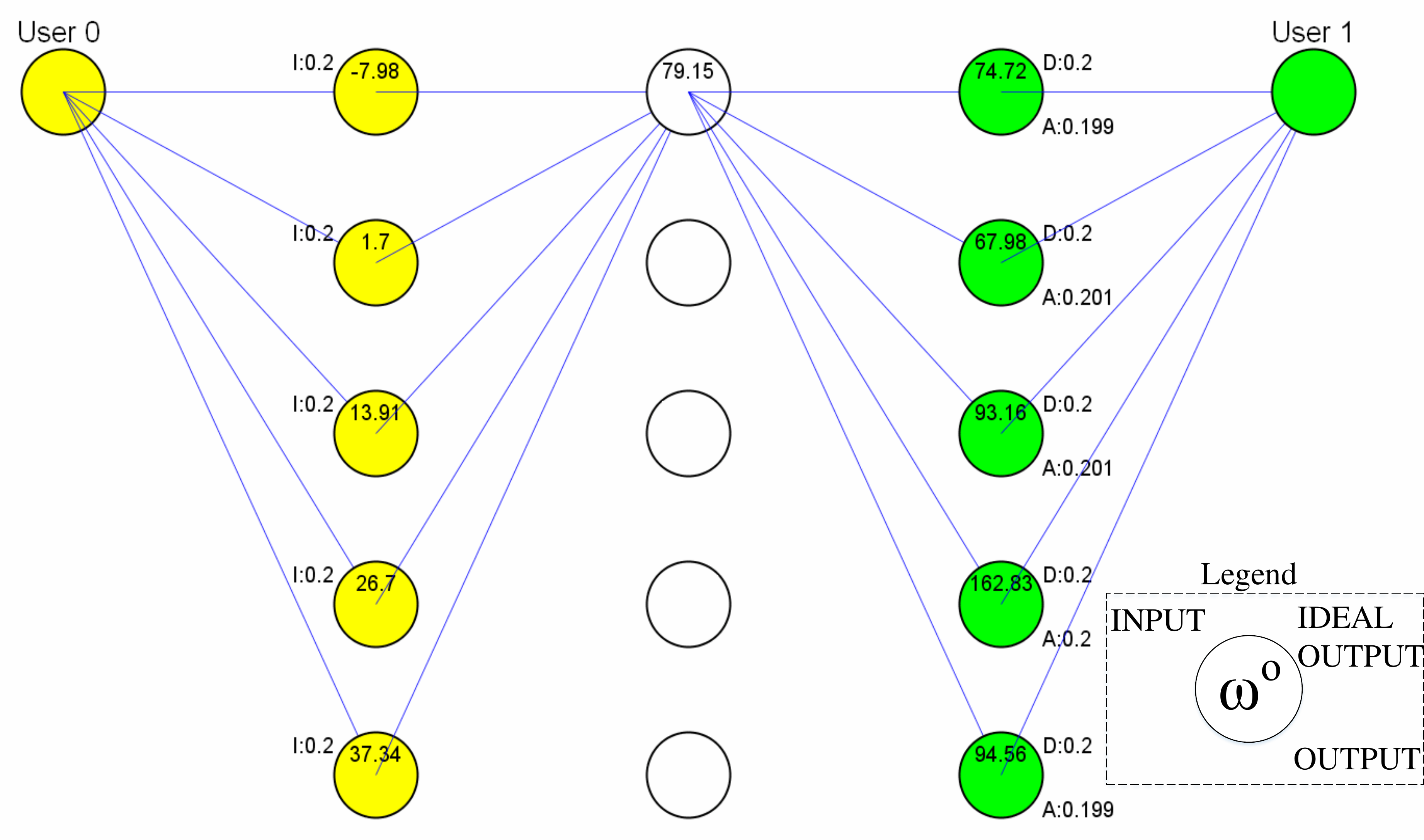}
\par\end{centering}
}
\par\end{centering}
\caption{\label{fig:NNTrain}The neural network corresponding to the studied
setup and its untrained/trained states. }
\end{figure}
 The \noun{NNConfig} neural network representation of the simulation
setup is shown in Fig.~\ref{fig:NNTrain}. The input to each tile
of the first layer is the normalized portion of power impinging upon
it via the links of user~$0$. Since the objective is to transfer
all emitted power to user~$1$, these inputs can be \emph{virtual}
(i.e., not equal to the actual impinging power distribution). Thus,
each input is set to $20$~\% of the total emitted power, for each
of the five tiles in the first layer. Using the same principle, the
ideal output is set to $20$~\% of the total power to reach the receiver.
Since the inputs and outputs remain the same at every feed-forward/back-propagate
cycle, a high learning rate is selected ($\eta=0.95$). The initial
$\omega$ angle values per tile are randomized in the range $\left[-90^{o},90^{o}\right]$.
This leads to the formation of the untrained network of Fig.~\ref{fig:Untrained}.
Finally, the termination criterion is to reach a root mean square
error (RMSE) of less that $10^{-3}$, which is achieved at approximately
$2000$ feed-forward/back-propagate cycles.

The trained network is shown in Fig.~\ref{fig:Trained}. Notably,
\noun{NNConfig }finds a solution where only one tile is activated
in the middle layer. This is the natural outcome of reinforcement
learning around an ``impactful'' tile, i.e., one already managing
a considerable part of the total impinging power. Thus, \noun{NNConfig
}is economic in terms of used tiles, which can be beneficial in terms
of power expenditure and supported user capacity.

\begin{figure}[t]
\begin{centering}
\subfloat[\label{fig:NatProp}Natural propagation. ]{\begin{centering}
\includegraphics[width=0.45\columnwidth,height=3.3cm]{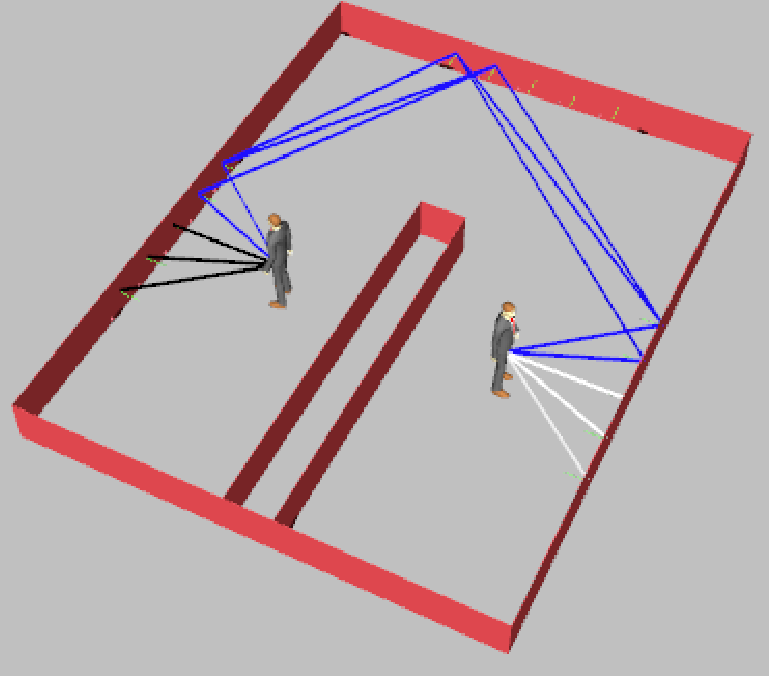}
\par\end{centering}
}\subfloat[\label{fig:KpProp}Propagation via $\text{\textsc{KPConfig}}$. ]{\begin{centering}
\includegraphics[width=0.45\columnwidth]{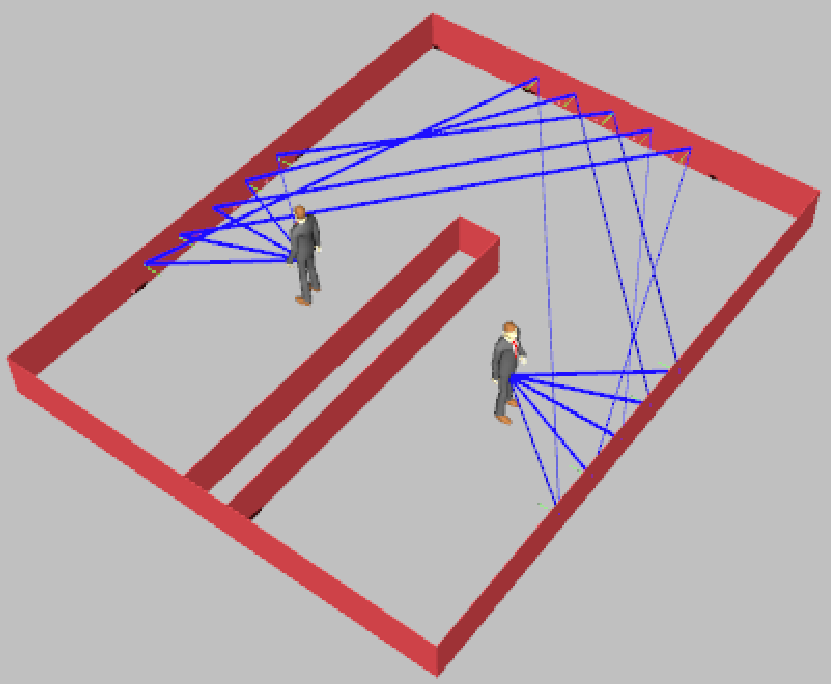}
\par\end{centering}
}
\par\end{centering}
\begin{centering}
\subfloat[\label{fig:nnconfig}Propagation via $\text{\textsc{NNConfig}}$. ]{\begin{centering}
\includegraphics[width=0.45\columnwidth,height=3.3cm]{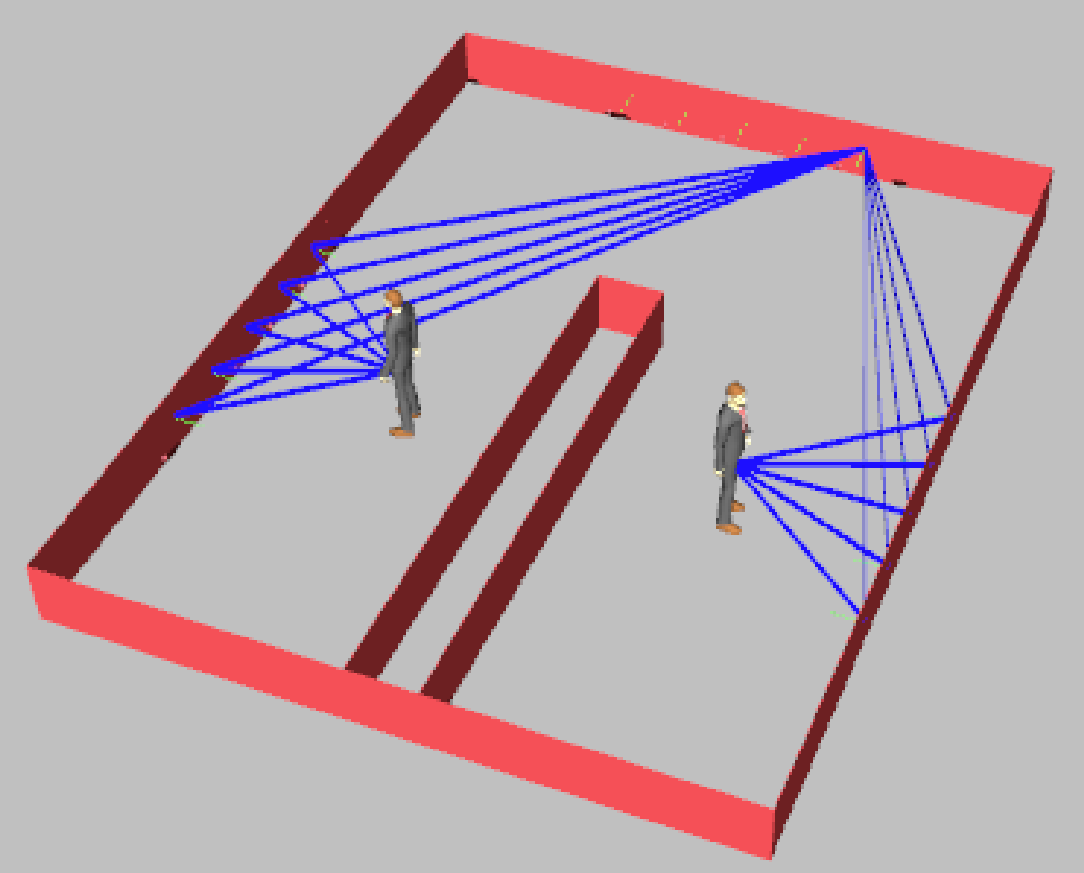}
\par\end{centering}
}
\par\end{centering}
\caption{\label{fig:Propagation}Wireless propagation visualized for each approach. }
\end{figure}
Visualizations of the ray-traced outcome per each compared scheme
are given in Fig.~\ref{fig:Propagation}. Regular propagation supports
only symmetric (specular) ray reflection. As a result, it cannot steer
$\sim60$~\% of the emitted power to the receiver; since 3 out of
5 rays (to the front and right of the transmitter) are directly reflected
to absorbing areas. \noun{KpConfig} works as intended and finds a
solution that routes all power to the receiver. However, it uses all
tiles in the middle wall, reaching full PWE capacity. \noun{NNConfig
}uses just one tile in the middle, in perfect match with the neural
network abstraction in Fig.~\ref{fig:Trained}.

\begin{table}[t]
\centering{}\caption{\textsc{\label{tab:RPow}Received Power per Approach.}}
\begin{tabular}{|c|c|}
\hline
Regular propagation & $-70.71$~$dBmW$\tabularnewline
\hline
$\text{\textsc{KPConfig}}$ & $-49.87$~$dBmW$\tabularnewline
\hline
$\text{\textsc{NNConfig}}$ & $-49.83$~$dBmW$\tabularnewline
\hline
\end{tabular}
\end{table}
The received power per scheme is given in Table~\ref{tab:RPow}.
All three propagation approaches are subject to receiving antenna
aperture losses~\cite{liaskos2018modeling}. Regular propagation
performs worse for two reasons. First, it loses $\sim60$~\% of the
emitted power for the reasons described above. Second, it does not
support collimation and focusing capabilities and is, thus, subject
to the $\nicefrac{1}{r^{2}}$ power attenuation rule~\cite{liaskos2018modeling}.
On the other hand, both \noun{KpConfig} and \noun{NNConfig }behave
similarly in terms of received power, since they successfully groom
and route all power from the transmitter to the receiver.

\subsubsection*{Discussion and future work}

\noun{NNConfig} exhibits potential for configuring PWEs in a tile-economic
manner. It can minimize the number of tiles required for serving communication
objectives. Moreover, it follows a representation of the PWE setting
that is simple, intuitive and directly interpretable. Future work
is directed towards extending the neural network representation to
multiple users and multiple objectives. To this end, a promising direction
is its combination with \noun{KpConfig} in a hierarchical approach,
where \noun{KpConfig} coarsely routes EM waves at the level of wall
orderings, and \noun{NNConfig} deduces the exact EM functionality
per tile.

\section{Conclusion\label{sec:Conclusion}}

Programmable wireless environments allow for software-defined propagation
of electromagnetic waves within a space, with tremendous potential
for wireless communication. Special tiles made from planar, software-defined
metamaterials receive programmatic commands and change their interaction
with impinging electromagnetic waves, benefiting wireless devices.
The present paper introduced an interpretable neural network-based
approach for configuring the behavior of tiles in such environments.
Tiles and inter-tile power flow were mapped to neural network nodes
and links respectively, and custom feed-forward and back-propagate
processes were introduced. Evaluation via ray-tracing showed performance
potential at the level of state-of-the-art solutions, with a distinct
gain in reducing the total number of tiles required to be activated.

\section*{Acknowledgment}

This work was funded by the European Union via the Horizon 2020: Future
Emerging Topics call (FETOPEN), grant EU736876, project VISORSURF
(http://www.visorsurf.eu).


\end{document}